\documentclass[journal,10pt,twocolumn]{IEEEtran}
\usepackage{cite}
\usepackage{amsmath,amssymb,amsfonts}
\usepackage{algorithmic}
\usepackage{graphicx}
\usepackage{textcomp}

\usepackage{subfig}
\usepackage{multirow}

\newtheorem{Definition}{Definition}
\newtheorem{Theorem}{Theorem}
\newtheorem{Lemma}{Lemma}
\newtheorem{Remark}{Remark}
\newtheorem{Problem}{Problem}
\newtheorem{Corollary}{Corollary}
\newtheorem{Assumption}{Assumption}

\usepackage{xcolor}
\usepackage{hyperref}
 \hypersetup{
   colorlinks  = true,
   urlcolor    = blue,
   linkcolor   = red, 
   citecolor   = red
 }

\usepackage{amsmath,bm}
\usepackage{amsmath,amssymb,mathrsfs}

\newcommand{\m}[1]{\mathbf{#1}}
\newcommand{\mc}[1]{\mathcal{#1}}
\newcommand{\mb}[1]{\mathbb{#1}}
\usepackage{savesym}
\usepackage{amsmath}

\usepackage{tikz}
\usetikzlibrary {arrows.meta,bending,positioning,shapes.geometric}

\begin{document}

\title{The networked input-output economic problem}
\author{Minh Hoang Trinh, Nhat-Minh Le-Phan, Hyo-Sung Ahn
\thanks{M. H. Trinh is with the AI Department, FPT University, Quy Nhon AI Campus, Quy Nhon, Binh Dinh, Vietnam (e-mail: minhtrinh@ieee.org).}
\thanks{N.-M. Le-Phan is with Department of Automation Engineering, School of Electrical and Electronic Engineering, Hanoi University of Science and Technology (HUST), Hanoi 11615, Vietnam. (e-mail: minh.lephannhat@hust.edu.vn).}
\thanks{School of Mechanical Engineering, Gwangju Institute of Science and Technology (GIST), Gwangju, Republic of Korea. (e-mail: hyosung@gist.ac.kr).}
}

\maketitle

\begin{abstract}
In this chapter, an input-output economic model with multiple interactive economic systems is considered. The model captures the multi-dimensional nature of the economic sectors or industries in each economic system, the interdependencies among industries within an economic system and across different economic systems, and the influence of demand. To determine the equilibrium price structure of the model, a matrix-weighted updating algorithm is proposed. The equilibrium price structure is proved to be globally asymptotically achieved when certain joint conditions on the matrix-weighted graph and the input-output matrices are satisfied. The theoretical results are then supported by numerical simulations. 
\end{abstract}


The hidden interdependent structure between different industries or economic sectors inside a national economic systems has been identified and eventually modeled by the economist W. Leontief, as early as from the 1930s \cite{Carvalho2019}. In the Leontief's input-output economic model, each industry is assumed to produce a single type of product by consuming the products from other industries. The relation between the outputs of different industrial sector and the demand is described by a linear balance equation, in which an input-output matrix (or input-output table) links the output with the demand. The solution of the balance equation is called the equilibrium price vector, which can be used as a tool for planning or proposing new policies to amend the current economic network \cite{Leontief1986input}.

This chapter first attempts to model a global economic network system based on the notion of matrix-weighted graphs \cite{Trinh2018matrix} and the work on inter-regional input-output model described by W. Isard \cite{Isard1966methods}[Chapter 8]. Each economic system (or an agent) is represented as a vertex in the matrix-weighted graph, and each edge in the graph is associated with a corresponding nonnegative matrix weight describing a piece of multidimensional trade relations between different industries within a country (self-loop) or between different countries. The matrix weighted graph provides a whole picture of the global input-output economic system. Second, the problem of determining the equilibrium price structure of the networked input-output economic system is considered for closed and open networks. In a closed model, the output of each industries is all consumed by some other industries. The characteristics of a closed networked economic system is thus fully determined by its corresponding aggregated input-output matrix. In contrast, in an open model, the production demands are also considered on the balancing equation relating the input-output production vectors. Distributed consensus-like algorithms are proposed to determine the solutions of the balancing equations in both models. Sufficient conditions are provided to guarantee the algorithms to asymptotically achieve the exact solution. Further, the updating algorithms also predict how the networked input-output systems would behave in the long term given a set of initial available goods. 

The remainder of this chapter is organized as follows. Section~\ref{sec:c15_modeling} introduces the theoretical foundations, including the Perron-Frobenius theorem and matrix-weighted graphs, and presents the formulation of Leontief's networked input-output model. Section~\ref{sec:c15_main} discusses distributed algorithms for computing the equilibrium output vectors in both closed and open versions of the model. Section~\ref{sec:c15_sim} demonstrates the theoretical results through numerical simulations. Finally, Section~\ref{sec:notes} concludes the chapter with a summary and potential directions for future work.
\section{Modeling}
\label{sec:c15_modeling}

\subsection{Perron-Frobenius theorem}
A matrix $\m{A}=[a_{ij}]\in\mb{R}^{n \times n}$, $n \in \mb{N}_+$, is nonnegative (positive), denoted $\m{A}\geq 0$ (resp., $\m{A}>0$), if it has all non-negative (resp., positive) entries, i.e., $a_{ij}\geq 0$ (resp., $a_{ij}>0$), $\forall i,j\in \{1,\ldots,n\}$. A nonnegative matrix $\m{A}$ is irreducible if and only if $\sum_{k=0}^{n-1}\m{A}^k>0$, and it is primitive if and only if there exists $k$ such that $\m{A}^k>0$. \index{irreducible matrix} \index{primitive matrix}

If $\m{A}$ is non-negative\index{non-negative matrix}\index{positive matrix} and $\m{A}^\top\m{1}_n = \m{1}_n$, where $\m{1}_n\in\mb{R}^n$ is the vector of all ones, then $\m{A}$ is column-stochastic. A matrix is column-substochastic if and only if $\m{A}^\top\m{1}_n \leq \m{1}_n$ and there exists a column sum which is strictly smaller than 1.\index{column-stochastic} \index{column sub-stochastic}

Let $G=(V,E,W)$ be a directed scalar-weighted graph with the vertex set $V=\{v_1,\ldots, v_n\}$, the edge set $E=\{e_1,\ldots,e_m\}\subseteq V\times V$ counting self-loops, and a weight set $W=\{w_{e_1},\ldots,w_{e_m}\}$ of positive edge weights. For each edge $e_k\equiv (v_j,v_i)$ from $v_j$ to $v_i \in V$, there is a corresponding positive edge weight $w_{e_k}\equiv w_{ij}>0$.

Let $\m{W}=[w_{ij}] \in \mb{R}^{n \times n}$ be the adjacency matrix  corresponding to the graph $G$, then $\m{W}$ is nonnegative. The graph $G$ is strongly connected if and only if $\m{W}$ is irreducible. Moreover, $G$ is a strongly connected aperiodic graph if and only if $\m{W}$ is primitive. 

The Perron-Frobenius theorem for column-stochastic adjacency matrices is summarized on the following lemma.\index{Perron-Frobenius theorem}

\begin{Lemma}\cite{Bullo2019lectures} \label{lem:c15_Perron-Frobenius} Let $\m{W}\in\mb{R}^{n \times n}$, $n \in \mb{N}_+$ be a non-negative matrix with a corresponding graph $G$.
\begin{itemize}
\item[i.] If $G$ is strongly connected, $\m{W}$ has a simple dominant eigenvalue $\mu_{\max}$, and all other eigenvalues $\mu$ satisfy $0\leq |\mu| \leq \mu_{\max}$, the left eigenvector $\m{p}$ and the right eigenvector $\bm{\gamma}$ corresponding to the dominant eigenvalue of $\m{W}$ are positive and unique up to a scale.
\item[ii.] If $G$ is strongly connected and $\m{W}$ is column-stochastic, then $\mu_{\max}=1$. If one further imposes that $G$ is aperiodic, then $|\mu|<1$ for all eigenvalues differing from 1, and $\m{W}^k$ is semi-convergent, i.e., \index{semi-convergence}
\begin{align} \label{eq:c15_convergence}
\lim_{k\to+\infty}\m{W}^k = \bm{\gamma}\m{1}^\top_n,
\end{align}
where $\bm{\gamma}>0$, $\bm{\gamma}^\top\m{1}_n=1$, is the unique normalized right-eigenvector of $\m{W}$ corresponding to the eigenvalue 1.
\item[iii.] If $G$ has an in-root and $\m{W}$ is column-stochastic, then $\m{W}$ has a unique normalized right-eigenvector $\bm{\gamma}=[\gamma_1,\ldots,\gamma_n]^\top \geq 0$, $\m{1}_n^\top\bm{\gamma} = 1$, with $\gamma_i>0$ for each $v_i$ belongs to the subgraph of all in-roots in $G$ and $\gamma_i=0$, otherwise. Moreover, 
\[\lim_{k\to+\infty}\m{W}^k = \bm{\gamma}\m{1}_n^\top.\]
\end{itemize}
\end{Lemma}

\subsection{Leontief's input-output economic model}
Consider an economic system with $d\ge 2$ interdependent industries, each of which produces a single good and uses some of the production (output) of the other industries \cite{Leontief1986input,NicholsonLAA}. Let $z_p$ denote the output from the $p$-th industry. To produce an unit of good, the industry $p$ must consume $a_{pq}\geq 0$ units from the sector $q\in \{1,\ldots,d\}$. 

Thus, the price of the $p$-th product comprises of the consumption charges for other industries $a_{pq}z_q$ and a demand $y_p \ge 0$ from the market. One may write the equation
\begin{align} \label{eq:c15_input-output-model}
z_p = \sum_{q=1}^d a_{pq}z_q + y_p,
\end{align}
where the coefficients satisfy $0\leq a_{pq}\leq 1,~q=1,\ldots, d$, and $\sum_{p=1}^d a_{pq}\leq 1,~\forall q=1,\ldots, d$. Let $\m{A}=[a_{pq}]\in\mb{R}^{d\times d}$, $\m{z}_i = [z_1,\ldots,z_d]^\top$ and $\m{y}=[y_1,\ldots,y_d]^\top \geq 0$ denote the input-output matrix, the output vector, and the demand vector, respectively. The equation~\eqref{eq:c15_input-output-model} can be expressed in the matrix form as follows
\begin{align} \label{eq:c15_input-output-model-matrix-form}
\m{z} = \m{A}\m{z} + \m{y}.
\end{align}
The input-output economic problem involves the determination of the output vector $\m{z}$.

In a \emph{closed} economic system\index{closed model}, the entire output of each industry is being used by itself and other industries and there is no demand. Thus, $\m{A}$ is column-stochastic, $\m{y} = \m{0}_d$, and the equation \eqref{eq:c15_input-output-model-matrix-form} reduces to $\m{z}=\m{A}\m{z}$. The solution of the equation, $\m{z}$, is a positive eigenvector corresponding to the unity eigenvalue of $\m{A}$, and is called an \emph{equilibrium price structure}.

In general, the economic is \emph{open} \index{open model}with a nonzero demand vector ${y}\neq {0}$. If the matrix $\m{I}_d-\m{A}$ is invertible, the corresponding solution is uniquely given by $\m{z}=(\m{I}_d-\m{A})^{-1}\m{y}$. In the context of economic analysis, people are interested in nonnegative matrices having nonnegative inverses. These matrices are called productive matrices.\index{productive matrix}
\begin{Definition}[Productive matrix]\label{def:c15_productive_matrix} The matrix $\m{A}\in\mb{R}^{d \times d}$ is productive if and only if $\m{A}\ge 0$ and there exists a positive vector $\m{r}\in\mb{R}^d$ such that $\m{r}-\m{A}\m{r}$ is positive.
\end{Definition}
The following lemma provides necessary and sufficient conditions for a matrix to be productive.
\begin{Lemma}[Hawkins-Simon condition] \cite{Hawkins1948,Michel1989cours} \label{lem:c15_productive_matrix} The nonnegative matrix $\m{M}\in \mb{R}^{N\times N}$, $N\in \mb{N}_{+}$ is productive if and only if 
\begin{itemize}
\item[i.] $\m{N} = \m{I}_N-\m{M}$ is invertible with a nonnegative inverse, or
\item[ii.] all successive leading principal minors of $\m{N}=[n_{ij}]$ are positive, i.e., $n_{11}>0,\left|\begin{array}{cc}
n_{11} & n_{12} \\ n_{21} & n_{22}
\end{array} \right|>0,\ldots,$ and
\begin{align*}
\left|\begin{array}{cccc}
n_{11} & n_{12} & \ldots & n_{1N}\\ 
n_{21} & n_{22} & \ldots & n_{2N}\\
\vdots & \vdots & \ddots & \vdots \\
n_{N1} & n_{N2} & \ldots & n_{NN}
\end{array} \right|>0.
\end{align*}
\end{itemize}
\end{Lemma}
\subsection{The networked input-output model and problem formulation}
\label{subsec:c15_consensus_model}
Consider an interconnected $n\geq 2$ economic systems (agents), where each economic system has $d$ interdependent industries. By interconnectedness, each agent can trade with each others. To capture the interconnected economic systems, a \emph{matrix-weighted graph} ${G}=({V},{E},{W})$ is used \cite{Trinh2018matrix}. Each vertex $v_i\in {V}$ represents an $i$-th economic system, while each edge $(v_j,v_i)\in {E}$ implies that the $i$-th economic system consumes at least one type of products from the $j$-th economic system. 

Each economic system may choose to consume the products from itself (domestic products), import the products from other economic system, or blend between two options. Let the set of matrix weights
\[{W}=\{w_{ij}\m{A}_{ij}=[w_{ij}a_{i,j}^{p,q}]\in \mb{R}^{d \times d}|(v_j,v_i)\in E\}\] 
capture the consumption pattern inside the network. The following assumptions on the matrix weights will be considered in this chapter.

\begin{Assumption}\label{assumption:c15_1} The matrix-weighted graph ${G}$ contains at least an in-root, and the induced subgraph of all in-roots of ${G}$ is strongly connected and aperiodic. The matrix $\m{W}=[w_{ij}] \in \mb{R}^{n \times n}$ is column-stochastic, i.e., $w_{ij}$ satisfies $w_{ij}\in (0,1],~\forall (v_j,v_i)\in E$, and $\sum_{i=1}^n w_{ij}=1$.
\end{Assumption}

\begin{Assumption}\label{assumption:c15_2} 
The matrices $\m{A}_{ij}$ are primitive for all $(v_j,v_i)\in {E}(G)$. In addition, all column sums of $\m{A}_{ij}$ do not exceed unity, i.e., $\sum_{p=1}^d a_{i,j}^{p,q} \leq 1,\forall q=1,\ldots,d$.
\end{Assumption}
\begin{figure}[t]
    \centering
	    \resizebox{.9\linewidth}{!}{
    \begin{tikzpicture}[
roundnode/.style={circle, draw=black, thick, minimum size=2mm,inner sep= 0.25mm, fill = black},
squarednode/.style={rectangle, draw=black, thick, minimum size=3.5mm,inner sep= 0.25mm},
]
    \node[fill=black!10!white, ellipse, minimum width = 1cm, minimum height = 3.15cm, align=center] (e1) at (0,-1) {};
    \node[fill=black!10!white, ellipse, minimum width = 1cm, minimum height = 3.15cm, align=center] (e2) at (4,0) {};
    \node[fill=black!10!white, ellipse, minimum width = 1cm, minimum height = 3.15cm, align=center] (e3) at (2,-4) {};
    \node[fill=black!10!white, ellipse, minimum width = 1cm, minimum height = 3.15cm, align=center] (e4) at (6,-4) {};
    \node[fill=black!10!white, ellipse, minimum width = 1cm, minimum height = 3.15cm, align=center] (e5) at (8,-1) {};
    \node[roundnode]   (u1a)   at  (0,0)  { };     %
    \node[roundnode]   (u1b)   at  (0,-1) { };     %
    \node[roundnode]   (u1c)   at  (0,-2) { };     %
    \node[roundnode]   (u2a)   at  (4,1)  { };    %
    \node[roundnode]   (u2b)   at  (4,0) { }; %
    \node[roundnode]   (u2c)   at  (4,-1) { };%
    \node[roundnode]   (u3a)   at  (2,-3) { };%
    \node[roundnode]   (u3b)   at  (2,-4) { };%
    \node[roundnode]   (u3c)   at  (2,-5) { };%
    \node[roundnode]   (u4a)   at  (6,-3) { };%
    \node[roundnode]   (u4b)   at  (6,-4) { };%
    \node[roundnode]   (u4c)   at  (6,-5) { };%
    \node[roundnode]   (u5a)   at  (8,0) { };%
    \node[roundnode]   (u5b)   at  (8,-1) { };%
    \node[roundnode]   (u5c)   at  (8,-2) { };%

    \node[roundnode, fill = red]   (ul1)   at  (-1.5,0)  { };     %
    \node[roundnode, fill = red]   (ul2)   at  (-1.5,-1) { };     %
    \node[roundnode, fill = red]   (ul3)   at  (9,0) { };  
    \node[roundnode, fill = red]   (ul4)   at  (9,-1) { };  
    
    \draw [draw = red, dotted, very thick,-{Stealth[length=2mm]}]
    (ul1) edge [bend right=0] (u1a)
    (ul2) edge [bend right=0] (u1b)
    (ul3) edge [bend right=0] (u5a)
    (ul4) edge [bend left=0] (u5b);
    
    \draw [draw = blue, thick,-{Stealth[length=2mm]}]
    (u1a) edge [bend left=45] (u2a)
    (u1a) edge [bend left=45] (u2b)
    (u1a) edge [bend left=45] (u2c)
    (u1b) edge [bend left=0]   (u2a)
    (u1b) edge [bend left=0]   (u2b)
    (u1b) edge [bend left=0]   (u2c)
    (u1c) edge [bend right=30]  (u2a)
    (u1c) edge [bend right=30]  (u2b)
    (u1c) edge [bend right=30]  (u2c)
    ; 
    \draw [draw = violet, thick,-{Stealth[length=2mm]}]
    (u2a) edge [bend right=50] (u3a)
    (u2a) edge [bend right=40] (u3b)
    (u2a) edge [bend right=30] (u3c)
    (u2b) edge [bend left=0]  (u3a)
    (u2b) edge [bend left=0]  (u3b)
    (u2b) edge [bend left=0]  (u3c)
    (u2c) edge [bend left=45] (u3a)
    (u2c) edge [bend left=40] (u3b)
    (u2c) edge [bend left=30] (u3c)
    ; 
    \draw [draw = green, thick,-{Stealth[length=2mm]}]
    (u3a) edge [bend right=20] (u1a)
    (u3a) edge [bend right=45] (u1b)
    (u3a) edge [bend right=55] (u1c)
    (u3b) edge [bend left=0] (u1a)
    (u3b) edge [bend left=0] (u1b)
    (u3b) edge [bend left=0] (u1c)
    (u3c) edge [bend left=70] (u1a)
    (u3c) edge [bend left=60] (u1b)
    (u3c) edge [bend left=45] (u1c)
    ; 
    \draw [draw = black, thick,-{Stealth[length=2mm]}]
    (u1b) edge [in=135, out=-135,looseness=13] (u1b)
    (u1c) edge [in=-150, out=-60,looseness=20] (u1c)
    (u1a) edge [bend left=0] (u1b)
    (u1a) edge [bend left=45] (u1c)
    (u1b) edge [bend left=60] (u1a)
    (u1b) edge [bend left=0] (u1c)
    (u1c) edge [bend left=90] (u1a)
    (u1c) edge [bend left=90] (u1b)
    ; 
    \draw [draw = orange, thick,-{Stealth[length=2mm]}]
    (u4a) edge [bend right=45] (u3a)
    (u4a) edge [bend right=45] (u3b)
    (u4a) edge [bend right=45] (u3c)
    (u4b) edge [bend left=0] (u3a)
    (u4b) edge [bend left=0] (u3b)
    (u4b) edge [bend left=0] (u3c)
    (u4c) edge [bend left=45] (u3a)
    (u4c) edge [bend left=45] (u3b)
    (u4c) edge [bend left=45] (u3c)
    ; 
    \draw [draw = magenta, thick,-{Stealth[length=2mm]}]
    (u5a) edge [bend right=15] (u4a)
    (u5a) edge [bend right=10] (u4b)
    (u5a) edge [bend right=5] (u4c)
    (u5b) edge [bend left=0] (u4a)
    (u5b) edge [bend left=0] (u4b)
    (u5b) edge [bend left=0] (u4c)
    (u5c) edge [bend left=65] (u4a)
    (u5c) edge [bend left=55] (u4b)
    (u5c) edge [bend left=35] (u4c)
    ; 
    \draw [draw = cyan, thick,-{Stealth[length=2mm]}]
    (u5a) edge [bend right=35] (u2a)
    (u5a) edge [bend right=35] (u2b)
    (u5a) edge [bend right=35] (u2c)
    (u5b) edge [bend left=0] (u2a)
    (u5b) edge [bend left=0] (u2b)
    (u5b) edge [bend left=0] (u2c)
    (u5c) edge [bend left=25] (u2a)
    (u5c) edge [bend left=25] (u2b)
    (u5c) edge [bend left=25] (u2c)
    ; 
    \node (agent1) at (-0.5,1) {\Large Agent $1$};
    \node (agent2) at (4,2) {\Large Agent $2$};
    \node (agent3) at (.5,-5) {\Large Agent $3$};
    \node (agent4) at (7.5,-5) {\Large Agent $4$};
    \node (agent5) at (8,1) {\Large Agent $5$};
    \node (leader1) at (-1.5,0.5) {L$1$};
    \node (leader2) at (-1.5,-0.5) {L$2$};
    \node (leader4) at (9.0,-0.5) {L$4$};
    \node (leader3) at (9,0.5) {L$3$};
\end{tikzpicture}}
    \caption{A network of five economic systems (agents). Each economic system maintains three industries described by three black nodes. These edges describe a consumption link or the influence of a market demand.}
    \label{fig:c15_network}
\end{figure}
Let the state vector $\m{x}_i$, $\m{y}_i \in \mb{R}^d$ denote the output and the demand vectors of the $i-$th economic system, respectively. One can write a system of networked input-output equations
\begin{align}\label{eq:c15_networked_leontiff_model}
\m{x}_i = \sum_{j \in \mc{N}_i} w_{ij}\m{A}_{ij}\m{x}_j + \m{y}_i,~\forall i=1,\ldots,n,
\end{align}
where $\mc{N}_i$ denotes the set of in-neighbor of vertex $v_i \in V$. Let $\m{x}=[\m{x}_1^\top,\ldots,\m{x}_n^\top]^\top$ and $\m{y}=[\m{y}_1^\top,\ldots,\m{y}_n^\top]^\top \in \mb{R}^{dn}$, the equation \eqref{eq:c15_networked_leontiff_model} can be expressed in the matrix form
\begin{align}\label{eq:c15_networked_leontiff_model_matrix}
\m{x} = \bar{\m{A}}\m{x} + \m{y},
\end{align}
where $\bar{\m{A}}=[w_{ij}\m{A}_{ij}] \in \mb{R}^{dn\times dn}$ is the \emph{matrix-weighted adjacency matrix}.

In this chapter, the following problem will be considered.
\begin{Problem}\label{c15_problem} Suppose that Assumptions~\ref{assumption:c15_1} and \ref{assumption:c15_2} hold. Design distributed algorithms to solve the networked input-output equation~\eqref{eq:c15_networked_leontiff_model_matrix} for 
\begin{itemize}
\item[i.] The closed model: the matrices $\m{A}_{ij}$, $(v_j,v_i)\in {E}$ are column-stochastic, and there is no demand $\m{y} = \m{0}_{dn}$.
\item[ii.] The open model: at least a matrix $\m{A}_{ij}=[a^{p,q}_{i,j}]$ corresponding to an edge $(v_j,v_i)$ of the induced subgraph ${H}$ of the in-roots in ${G}$ is column-substochastic; and there is a nonzero demand $\m{y} \ne \m{0}_{dn}$.
\end{itemize}
\end{Problem}

Figure~\ref{fig:c15_network} provides an example of a five-agent system, in which each agent has three basic industries. In case there is no demand (closed model), one may omit the vertices L1--L4, and observe that agents 1, 2, and 3 belong to the set of the in-roots of the matrix-weighted graph ${G}$. In an open model, demands are non-zero at the first- and the second layers of agent 1, and the first- and the second layer of agent 5.

\section{Distributed computation of the equilibrium output vector}
\label{sec:c15_main}
In this section, a distributed algorithm, which has form of a matrix-weighted consensus algorithm, for solving the Problem~\ref{c15_problem} is proposed. Then, the convergence analysis for the closed and open networked input-output models under the proposed algorithm is provided.

\subsection{The proposed algorithm}
Problem~\ref{c15_problem} requires solving a set of linear equations, with column-substochastic matrix weights $w_{ij}\m{A}_{ij}$. 

Each agent $i\in V$ has a state vector $\m{x}_i[k] \in \mb{R}^d$ representing the outputs of the $i$-th system, and it knows the state vector of incomming neighbors in the network. At each discrete-time instant, agent $i$ updates its state vector based on the following updating algorithm
\begin{align}\label{eq:c15_dmwc}
\m{x}_i[k+1] = \sum_{j \in \mc{N}_i}w_{ij}\m{A}_{ij}\m{x}_j[k] + \m{y}_i,~k=0,1,2,\ldots
\end{align}

The networked input-output system can be considered as a multi-layer system. Each agent has $d$ states corresponding to $d$ layers, each layer corresponds to an economic sector. The self-loop matrix weight $\m{A}_{ii}=[a_{ii}^{pq}]$ represents the input-output interdependencies inside the agent $i$ (interlayer interactions). When agent $i$ is isolated from other agents, the balance equation written for agent $i$ is exactly the Leontief's input-output model \eqref{eq:c15_networked_leontiff_model_matrix}.

In each matrix 
\begin{align*}
    \m{A}_{ij} = [a_{i,j}^{p,q}]_{d\times d} =
    \begin{bmatrix}
        a_{i,j}^{1,1} & a_{i,j}^{1,2} & \ldots & a_{i,j}^{1,d}\\
        a_{i,j}^{2,1} & a_{i,j}^{2,2} & \ldots & a_{i,j}^{2,d}\\
        \vdots & \vdots & \ddots  & \vdots \\
        a_{i,j}^{d,1} & a_{i,j}^{d,2} & \ldots & a_{i,j}^{d,d}
    \end{bmatrix},~ i\neq j,
\end{align*}
the element $a_{i,j}^{p,q}$ represents the consumption of products from the $q-$th industry of the economic system $j$ by the $p$-th industry of the economic system $i$. Then, $a_{i,j}^{p,p}$ represents an intra-layer interaction and $a_{i,j}^{p,q},~p\neq q$ represents a cross-layer interaction between two economic systems $j$ and $i$. These inter-dependencies can be visualized by explicitly writing Eq.~\eqref{eq:c15_dmwc} as follows \cite{Ahn2020opinion}
\begin{align} \label{eq:c15_consensus}
  {x}_{ip}[k+1] &= \underbrace{\sum_{q=1}^d w_{ii}a_{i,i}^{p,q}x_{iq}[k]}_{\text{inter-layer interactions}} + \underbrace{\sum_{\substack{j=1,\\ j\neq i}}^n w_{ij}a_{i,j}^{p,p}x_{jp}[k]}_{\text{intra-layer interactions}} \nonumber\\
  &\quad + \underbrace{\sum_{\substack{j=1,\\j\neq i}}^n\sum_{\substack{q=1,\\ q\neq p}}^d w_{ij}a_{i,j}^{p,q}x_{jq}[k]}_{\text{cross-layer interactions}} + y_{ip},
\end{align}
for all $i=1,\ldots,n$, $p=1,\ldots,d$ and $k=0,1,2,\ldots$.

The linear recurrence equation~\eqref{eq:c15_dmwc} defines a trajectory of the networked input-output systems \eqref{eq:c15_input-output-model-matrix-form} associated with a given initial output vector that leads to a solution of \eqref{eq:c15_input-output-model-matrix-form}. If Eq.~\eqref{eq:c15_input-output-model-matrix-form} has a forward recursive solution, it means that the equation represents a self-determined dynamic economy, of which the next output level can be determined from the input of the current output and the demand vector \cite{Luenberger1977singular}. 

\subsection{The closed networked input-output model}
In this subsection, the closed model is considered. Let ${S}$ be the set of in-roots in ${G}$. Without loss of generality, suppose that the vertices belonging to ${S}$ are labeled $v_1$, $\ldots$, $v_{|S|}$. 

\begin{Lemma}\label{lem:c15_closed_col_stochastic}
Under the Assumptions~\ref{assumption:c15_1}--\ref{assumption:c15_2}, the matrix $\bar{\m{A}}$ corresponding to the closed input-output model \eqref{eq:c15_networked_leontiff_model_matrix} is column-stochastic.
\end{Lemma}

\begin{IEEEproof}
The $i$-th column sum of the matrix $\bar{\m{A}}$ is given by
\begin{align*}
\sum_{i=1}^n\sum_{p=1}^d w_{ij} a_{i,j}^{p,q} &= \sum_{i=1}^n w_{ij}\left(\sum_{p=1}^d a_{i,j}^{p,q}\right)=\sum_{i=1}^n w_{ij} = 1,
\end{align*}
for all $i=1,\ldots,dn$. It follows that $\bar{\m{A}}$ is column-stochastic.
\end{IEEEproof}

\begin{Lemma}\label{lem:c15_closed_col_in_roots} Consider the closed input-output model under the Assumptions~\ref{assumption:c15_1}--\ref{assumption:c15_2}. Let $\bar{G}=(\bar{V},\bar{E},\bar{A})$ be the graph corresponding to the matrix $\bar{\m{A}}$, i.e., $\bar{V}=\{\bar{v}_1,\ldots,\bar{v}_{dn}\}$, $\bar{E}\subseteq \bar{V}\times \bar{V}$, $\bar{A}=\{\bar{a}_{(i-1)d+p,(j-1)d+q} = w_{ij}a^{p,q}_{i,j}|~({v}_j,{v}_i)\in {E}, p, q=1,\ldots,d\}$. Then $\bar{G}$ contains a set of in-roots $\bar{v}_{(i-1)d+q}$, where $i=1,\ldots, |{S}|$ and $q=1,\ldots,d$. The induced subgraph of the in-roots in $\bar{G}$ is strongly connected and aperiodic.
\end{Lemma}

\begin{IEEEproof}
As the graph ${G}$ contains a subgraph of $|{S}|$ in-roots, each vertex in ${S}$ can be reached from any vertex $v_i$, $i=1,\ldots,n$. 

Consider the partition $\bar{V}=\bigcup_{k=1}^n\bar{V}_k$, $\bar{V}_i \cap \bar{V}_j = \emptyset,~\forall i\neq j$, where each set $\bar{V}_k=\{\bar{v}_{(k-1)d+1},\ldots,\bar{v}_{kd}\}$ corresponds to a vertex $v_k \in {G}$. As the graph $\bar{G}$ corresponds to the matrix $\bar{\m{A}}$, it follows that each set $\bar{V}_i,~i=1,\ldots,|{S}|$ can be reached by any $\bar{V}_k, k=1,\ldots, n$.

Consider a vertex $\bar{v}_l \in \bar{G}$ such that $1 \leq \lceil\frac{l}{d}\rceil \leq |S|$. One shows that $\bar{v}_l$ can be reached by any vertex $\bar{v}_k \in \bar{G}$. 
\begin{itemize}
\item Case 1: $\lceil\frac{l}{d}\rceil = \lceil\frac{k}{d}\rceil$. The vertex ${v}_{\lceil\frac{k}{d}\rceil} \in {S}$. Since $\m{A}_{ij}$ is primitive, it follows from Lemma~\ref{lem:c15_Perron-Frobenius} that the graph corresponding to each matrix $\m{A}_{ij}$ is strongly connected and aperiodic. In the graph $\bar{G}$, the vertex $\bar{v}_l$ can be reached from any other vertices $\bar{v}_k \in {V}_{\lceil\frac{l}{d}\rceil} = \{\bar{v}_{(\lceil\frac{l}{d}\rceil-1)d+1}, \ldots, \bar{v}_{\lceil\frac{l}{d}\rceil d}\}$. Reversely, there is also a directed path from $\bar{v}_l$ to $\bar{v}_k$. 
\item Case 2: Consider an arbitrary vertex $\bar{v}_k \notin \bar{V}_{\lceil\frac{l}{d}\rceil}$, then $\bar{v}_k$ belongs to some $\bar{V}_{\lceil\frac{k}{d}\rceil}$. Thus, there exists a directed path from some vertex in $\bar{V}_{\lceil\frac{k}{d}\rceil}$ to some vertex in $\bar{V}_{\lceil\frac{l}{d}\rceil}$. Let $\bar{V}_{k_1},\ldots,\bar{V}_{k_t}$ be a sequence of vertex set corresponding to such a path, with $k_1\equiv \lceil\frac{k}{d}\rceil$ and $k_t\equiv \lceil\frac{l}{d}\rceil$. Then, one may construct a directed path $\bar{\mc{P}} \in \bar{G}$ starting from $\bar{v}_k$ to $\bar{v}_l$ containing the following paths:
\begin{itemize}
\item $\bar{\mc{P}}_1$ starts from $\bar{v}_k$ to some vertex $\bar{v}_{k_2}$ in $\bar{V}_{k_2}$, where $\bar{v}_{k_2}$ is the out-neighbor of some vertex $\bar{v}_{k_1}' \in \bar{V}_{\lceil\frac{k}{d}\rceil}$.
\item $\bar{\mc{P}}_2$ starts from $\bar{v}_{k_2}$ to some vertex $\bar{v}_{k_3}$ in $\bar{V}_{k_3}$, where $\bar{v}_{k_3}$ is the out-neighbor of some vertex $\bar{v}_{k_2}' \in \bar{V}_{k_2}$.
\item $\ldots$
\item $\bar{\mc{P}}_{t-1}$ starts from $\bar{v}_{k_{t-1}}$ to some vertex $\bar{v}_{k_t}$ in $\bar{V}_{k_t}\equiv V_{\lceil\frac{l}{d}\rceil}$, where $\bar{v}_{k_t}$ is the out-neighbor of some vertex $\bar{v}_{k_{t-1}}' \in \bar{V}_{k_{t-1}}$.
\item $\bar{\mc{P}}_{t}$ starts from $\bar{v}_{k_{t}}$ to $\bar{v}_l$.
\end{itemize}
The existence of each path $\bar{\mc{P}}_{k_i},~i=1,\ldots,t$ follows from the assumption that the graph corresponding to each $\m{A}_{ij}$ is strongly connected.
\end{itemize}
As $\bar{v}_l$ can be chosen arbitrarily in $1 \leq \lceil\frac{l}{d}\rceil \leq |S|$, the set of vertices $\bar{S}=\{\bar{v}_l|1\leq l \leq d|S|\}$ forms a set of in-roots of $\bar{G}$.

Recall that the period of a strongly connected subgraph is the smallest common divisor of all cycles inside itself. Thus, if a strongly connected graph contains a strongly connected subgraph which is aperiodic, then it must be also aperiodic. Now, as the subgraphs of vertices in $\bar{V}_k$, $1\leq k\leq |S|$ are aperiodic, and belong to the induced subgraph corresponding to $\bar{S}$, one concludes that the induced subgraph corresponding to $\bar{S}$ is aperiodic. 
\end{IEEEproof}

Let $\m{x}[k] = [\m{x}_1[k]^\top,\ldots,\m{x}_n[k]^\top]^\top$, the closed networked input-output model can be expressed in the following matrix form
\begin{align}\label{eq:c15_mwc_closed}
\m{x}[k+1]=\bar{\m{A}}\m{x}[k],\quad \m{x}[0]=\m{x}_0
\end{align}

\begin{Theorem}\label{thm:c15_closed_IO_model}
Consider the closed input-output model with the Assumptions~\ref{assumption:c15_1}--\ref{assumption:c15_2}. Under the algorithm \eqref{eq:c15_dmwc},
\begin{align}
\lim_{k \to +\infty} \m{x}[k] = \bar{\bm{\gamma}}\m{1}_{dn}^\top\m{x}[0],
\end{align}
where $\bar{\bm{\gamma}} = [\bar{\gamma}_1,\ldots\bar{\gamma}_{dn}]^\top\geq 0$ is the right eigenvector corresponding to the eigenvalue 1 of $\bar{\m{A}}$, $\bar{\gamma}_i>0$ for $i=1,\ldots, d|{S}|$, $\sum_{i=1}^{d|{S}|}\bar{\gamma}_i = 1$. In other words, the system \eqref{eq:c15_dmwc} converges to an equilibrium price structure, which is only determined by the economic systems $1,\ldots,|{S}|$.
\end{Theorem}

\begin{IEEEproof}
It follows from Lemmas~\ref{lem:c15_closed_col_stochastic}--\ref{lem:c15_closed_col_in_roots} that the matrix $\bar{\m{A}}$ is column-stochastic and semi-convergent. Matrix $\bar{\m{A}}^\top$ has a unique normalized left eigenvector $\bar{\bm{\gamma}} = [\bar{\gamma}_1,\ldots\bar{\gamma}_{dn}]^\top\geq 0$ corresponding to the eigenvalue 1. Further, $\bar{\gamma}_i>0$ for $i=1,\ldots, d|{S}|$, $\sum_{i=1}^{d|{S}|}\bar{\gamma}_i = 1$. Thus, based on Lemma~\ref{lem:c15_Perron-Frobenius} (iii), 
\begin{align*}
\lim_{k\to+\infty}\m{x}[k] = \lim_{k\to+\infty} \bar{\m{A}}^k \m{x}[0] = \bar{\bm{\gamma}}\m{1}_{dn}^\top\m{x}[0].
\end{align*}
\end{IEEEproof}

In a closed input-output economic model, starting from an initial price structure $\m{x}_i[0],~i=1,\ldots,n$, let each economic system follow the producing pattern described by the matrix-weighted graph ${G}$ and the update law \eqref{eq:c15_dmwc}. Theorem~\ref{thm:c15_closed_IO_model} suggests that the economic systems belong to the set of in-roots ${S}$ eventually decide the equilibrium price structure. The output vectors of all economic systems outside the set of in-roots eventually vanish $\m{x}_i[k] \to \m{0}_d,~i\notin {S}$, as they are being uni-directionally consumed by the economic systems belong to the set of in-roots. The contributions of each economic system $i \in {S}$ in equilibrium are determined by the coefficients $\bar{\gamma}_{(\lceil\frac{i}{d}\rceil-1)d+1},\ldots,\bar{\gamma}_{\lceil\frac{i}{d}\rceil d}$. 

Another notable observation is that, the theorem allows some economic systems $i\in {S}$ to have $\m{A}_{ii}=\bm{\Theta}_d$. Thus, the production of economic system $i$ relies completely on the outputs of their out-neighbors, and their outputs are also consumed only by their in-neighbors. 

\begin{Corollary}[Network of identical sub-input-output matrices]\label{cor:c15_1} Assume that all economic systems in a closed input-output model have a same input-output matrix $\m{A}_{ij}=\m{A}\in\mb{R}^{d \times d}$, $\forall (v_j,v_i)$. Then, under the matrix-weighted updating algorithm \eqref{eq:c15_dmwc},
\begin{align*}
\lim_{k\to+\infty}\m{x}[k] = (\bm{\omega}\otimes\bm{\gamma})\m{1}_{dn}^\top\m{x}[0],
\end{align*}
where $\bm{\omega}=[\omega_1,\ldots,\omega_n]^\top \ge 0$ is the unique right eigenvector corresponding to the eigenvalue 1 of $\m{W}$, $\omega_i>0,\,\forall i \in {S}$, $\bm{\gamma}=[\gamma_1,\ldots,\gamma_d]^\top>0$ is the unique normalized eigenvector corresponding to the eigenvalue 1 of $\m{A}$, and `$\otimes$' denotes the Kronecker product.
\end{Corollary}

\begin{IEEEproof}
It is not hard to see that in this case $\bar{\m{A}}=\m{W}\otimes\m{A}$, which has the right Perron eigenvector $\bm{\omega}\otimes\bm{\gamma}$ corresponding to the unity eigenvalue.
\end{IEEEproof}

In a networked input-output system with the same matrix $\m{A}_{ij}$, there holds $\lim_{k \to +\infty} \m{x}_i[k] = \omega_i (\m{1}_{dn}^\top\m{x}[0]) \bm{\gamma}$, where $\omega_i$ is only dependent on $\m{W}$. The coefficients $w_{ij}$, thus, determine the influence of each economic system to the equilibrium price structure.

\begin{Remark}[Pagerank algorithm]
The updating algorithm \eqref{eq:c15_dmwc} with $\m{y}_i=\m{0}_d, \forall i=1,\ldots,d,$ is related with the Google PageRank algorithm \cite{Ishii2014pagerank}. The objective of the PageRank problem is to determine the importance of each web-page, which is modeled as a vertex (an agent) in a scalar-weighted graph (an abstraction of the internet). The importance of each webpage $i$ is the entry $\gamma_i$ in the Perron eigenvector $\bm{\gamma}$ of the corresponding adjacency matrix of the graph. \index{Pagerank algorithm}

The equation \eqref{eq:c15_dmwc} offers a block-matrix update algorithm for the PageRank problem by considering a set of $d$ related websites as an agent, and solving the Perron eigenvector $\bar{\bm{\gamma}}$ of the adjacency matrix $\bar{\m{A}}$ in a distributed, iterative manner. 

Since the hyperlinks between webpages are often highly clustering and the internet has abundant dangling vertices (vertices having no outgoing edges), Assumption \ref{assumption:c15_2} usually does not hold. To address this issue, a modified algorithm for estimating the eigenvector $\bar{\bm{\gamma}}$ is replacing $\m{A}_{ij}$ by ${\m{B}}_{ij} = (1-m)\m{A}_{ij}+\frac{m}{d}\m{I}_d$ for some small $m \in (0,1)$. As $\m{B}_{ij}$ is now a positive matrix, Assumption~\ref{assumption:c15_2} is obviously satisfied. Correspondingly, $\bar{\m{A}}$ is replaced by $\m{B} = [\m{B}_{ij}]$ in \eqref{eq:c15_mwc_closed}. The Perron eigenvector of the column stochastic matrix $\m{B}$ is guaranteed to exist, given that Assumption~\ref{assumption:c15_1} holds.
\end{Remark}

\subsection{The open networked input-output model}
In this subsection, the open networked input-output model under the matrix-weighted updating algorithm~\eqref{eq:c15_input-output-model} will be examined. The following lemma on the matrix $\bar{\m{A}}$ is firstly proved.

\begin{Lemma} \label{lem:c15_invertibility} Consider the open networked input-output model and suppose that Assumptions~\ref{assumption:c15_1}--\ref{assumption:c15_2} are satisfied. Then, the matrix $\m{I}_{dn}-\bar{\m{A}}$ is invertible.
\end{Lemma}

\begin{IEEEproof}
Since $\m{A}_{ij}$ are primitive, it cannot have a zero column. Combining with Assumption~\ref{assumption:c15_1}, it follows that $0<\sum_{p=1}^d a_{i,j}^{p,q}\leq 1$, for all $q=1,\ldots,d$. Thus, there exists a non-negative matrix $\m{E}_{ij}$ such that $\m{A}_{ij}+\m{E}_{ij}$ is column stochastic. Since there is a column-substochastic matrix among $\{\m{A}_{ij}\}_{(v_j,v_i) \in E(H)}$, there exist $i,j\in \{1,\ldots,n\}$ such that $\m{E}_{ij}\ne \bm{\Theta}_d$. Define the matrix $\bar{\m{E}}=[\m{E}_{ij}]\ge 0$, it follows that $\bar{\m{E}}\ne \bm{\Theta}_{dn}$ and the matrix $\bar{\m{A}}+\bar{\m{E}}$ is column stochastic. Moreover, for each $j=1,\ldots,n$,
\begin{align*}
\m{1}_{d}^\top \sum_{i=1}^n w_{ij}(\m{A}_{ij} + \m{E}_{ij}) &= \sum_{i=1}^n w_{ij}\m{1}_{d}^\top(\m{A}_{ij} + \m{E}_{ij}) \\
&= \sum_{i=1}^n w_{ij}\m{1}_{d}^\top.
\end{align*}
Hence, one can write
\begin{align*}
(\bar{\m{A}}+\bar{\m{E}})^\top\m{1}_{dn} &= \m{1}_{dn}.
\end{align*}
Since at least one column sum of $\bar{\m{A}}$ does not exceed 1, $\rho(\bar{\m{A}})\leq 1$. Based on Lemma~\ref{lem:c15_closed_col_in_roots}, the graph $\bar{G}$ has a subgraph of $|{S}|$ in-roots. If $\bar{\m{A}}$ has a unity eigenvalue, let $\m{r}$ be the Perron vector $\m{r} \ge 0$ of $\bar{\m{A}}$ that satisfies $r_1,\ldots,r_{d|{S}|}>0$ (elements corresponding to the in-roots) and $\bar{\m{A}}\m{r}=\m{r}$. Then
$1=\m{1}_{dn}^\top \m{r} = \m{1}_{dn}^\top(\bar{\m{A}}+\bar{\m{E}})\m{r}= \m{1}_{dn}^\top(\bar{\m{A}}\m{r})+\m{1}_{dn}^\top\bar{\m{E}}\m{r}=1+\m{1}_{dn}^\top\bar{\m{E}}\m{r} > 1$, where the inequality is strict due to the assumption that at least an $\m{A}_{ij}$ belonging to ${H}$ is column-substochastic. This contradiction implies that $\rho(\bar{\m{A}})<1$. Thus $\m{I}_{dn}-\bar{\m{A}}$ is invertible.
\end{IEEEproof}

\begin{Lemma} \label{lem:c15_barA_productive} Consider an open networked input-output system and suppose that Assumptions~\ref{assumption:c15_1}--\ref{assumption:c15_2} hold. In addition, suppose that for each $i =1,\ldots,n$, there exists a matrix weight $\m{A}_{ij}$ with $\sum_{p=1}^da_{i,j}^{p,q}<1,~\forall q=1,\ldots, d$. Then, the matrix $\bar{\m{A}}$ is productive.
\end{Lemma}
\begin{IEEEproof}
Similar to the proof of the Lemma~\ref{lem:c15_invertibility}, one has
\begin{align} \label{eq:c15_ineq_productive}
(\bar{\m{A}}+\bar{\m{E}})^\top\m{1}_{dn} = \m{1}_{dn}
\end{align}
The assumption on the matrix weights guarantees the existence of a positive entry in each column of $\bar{\m{E}}$. It follows from equation \eqref{eq:c15_ineq_productive} that $\m{1}_{dn} - \bar{\m{A}}^\top\m{1}_{dn} = \bar{\m{E}}^\top\m{1}_{dn}>0$. Thus, based on Lemma~\ref{lem:c15_productive_matrix}, the matrix $\bar{\m{A}}^\top$ is productive. As $(\m{I}_{dn}-\bar{\m{A}}^\top)^{-1} = ((\m{I}_{dn}-\bar{\m{A}})^{-1})^\top \ge 0$, it follows that the matrix $\bar{\m{A}}$ is productive.
\end{IEEEproof}

The asymptotic behavior of the open networked input-output system under the matrix-weighted update law~\eqref{eq:c15_dmwc} is summarized in the following theorem.
\begin{Theorem}\label{thm:c15_open-io-model} Consider an open networked input-output system and suppose that Assumptions~\ref{assumption:c15_1}--\ref{assumption:c15_2} are satisfied. Under the matrix-weighted update algorithm~\eqref{eq:c15_dmwc},
\begin{itemize}
\item[i.] The state vector $\m{x}[k]$ globally exponentially converges to the equilibrium price structure $\m{x}^{*} = \lim_{k\to+\infty}\m{x}[k] = (\m{I}_{dn}-\bar{\m{A}})^{-1}\m{y}$;
\item[ii.] Furthermore, if for each $i \in {V}$, there exists a matrix weight $\m{A}_{ij}$ with $\sum_{p=1}^da_{i,j}^{p,q}<1,~\forall q=1,\ldots, d$, then $\m{x}^{*} \geq 0$.
\end{itemize}
\end{Theorem}
\begin{IEEEproof}
Based on Lemma~\ref{lem:c15_invertibility}, the matrix $\bar{\m{A}}$ is invertible in both cases (i) and (ii). Thus, if the system~\eqref{eq:c15_dmwc} has a finite limit $\m{x}^{*}$, it must satisfy
\begin{align*}
\m{x}^{*} = \bar{\m{A}}\m{x}^{*}+\m{y} \text{ or equivalently, } \m{x}^* = (\m{I}_{dn}-\bar{\m{A}})^{-1}\m{y}.
\end{align*}
Let $\m{v}[k]=\m{x}[k]-\m{x}^*$, then
\begin{align}
\m{v}[k+1] &=\bar{\m{A}}\m{x}[k]+\m{y}-\m{x}^* \nonumber\\
&=\bar{\m{A}}\m{x}[k]+(\m{I}_{dn}-\bar{\m{A}})\m{x}^*-\m{x}^*\nonumber\\
&=\bar{\m{A}}\m{v}[k]. \label{eq:c15_v_equation}
\end{align}
It follows from $\rho(\bar{\m{A}})<1$ that $\m{v}[k]$ converges to $\m{0}_{dn}$ at geometric rate. Equivalently, $\m{x}[k]\to \m{x}^*$, as $k \to +\infty$. The claim in (ii) follows from the fact that $\bar{\m{A}}$ is a productive matrix, which has been shown in Lemma~\ref{lem:c15_barA_productive}.
\end{IEEEproof}

The condition imposing for $\bar{\m{A}}$ to be productive in Lemma~\ref{lem:c15_barA_productive} is only sufficient. Noting that a stricter condition to ensure $\bar{\m{A}}$ productive is for each $k=1,\ldots,dn$, there exists a matrix $\m{A}_{\lceil\frac{k}{d}\rceil j}$ such that $\sum_{p=1}^d a_{\lceil\frac{k}{d}\rceil, j}^{p,k-\lfloor\frac{k}{d}\rfloor}<1$. However, this condition does not offer a clear economic interpretation, unlike the condition stated in Lemma~\ref{lem:c15_barA_productive} and Theorem~\ref{thm:c15_open-io-model} (ii).

\begin{Remark}
The problem of determining the fittest solution to a set of linear equations $\m{A}\m{x}=\m{b}$ in a distributed manner was considered in \cite{Mou2015distributed,Shi2016network,Pham2023distributed}. There are mainly two approaches namely the iteration-based and the optimization based algorithms. The algorithm \eqref{eq:c15_dmwc} is iteration-based, which guarantees convergence to one of the exact solution of the equation. The optimization based method in \cite{Pham2023distributed} can also be applied, which gives a least square solution of the problem in all cases. The algorithm~\eqref{eq:c15_dmwc} and its convergence requirements are presented since they intuitively describe how the networked input-output system evolves according to a given input-output matrix and demand vector.
\end{Remark}

\section{Simulation results}
\label{sec:c15_sim}
In this section, simulation examples are provided to illustrate the theoretical results in Sections~\ref{sec:c15_main}. A network of five primitive societies as depicted in Fig.~\ref{fig:c15_network} is considered.
\begin{figure}[t]
    \centering
    \subfloat[]{\includegraphics[width=0.45 \linewidth]{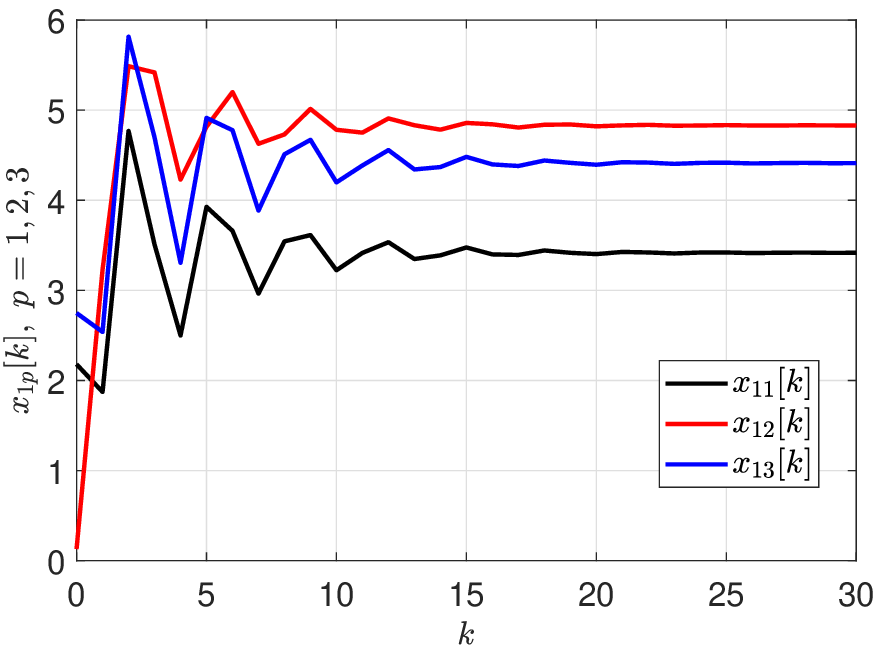}} \hfill
    \subfloat[]{\includegraphics[width=0.45 \linewidth]{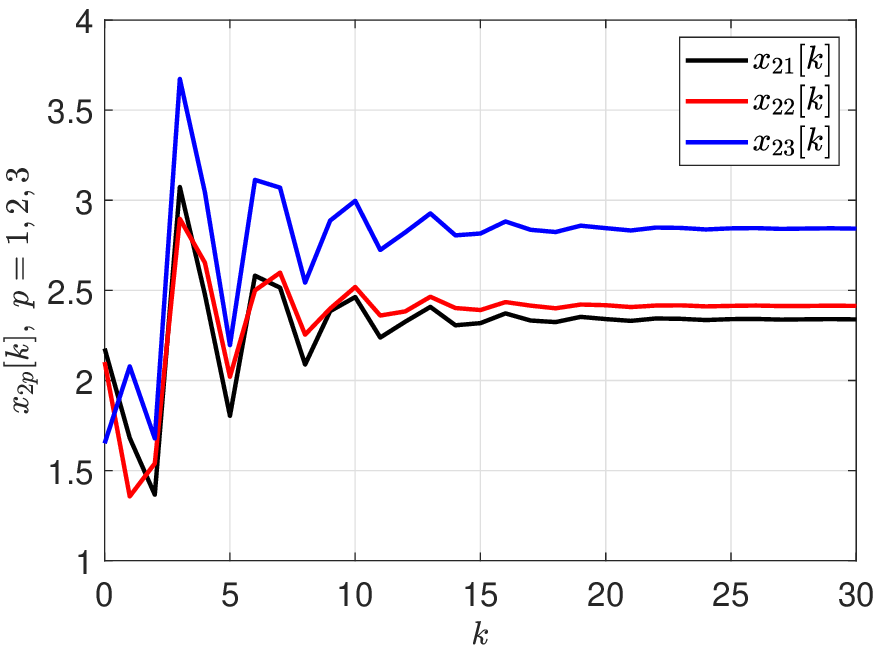}} \hfill
    \subfloat[]{\includegraphics[width=0.45 \linewidth]{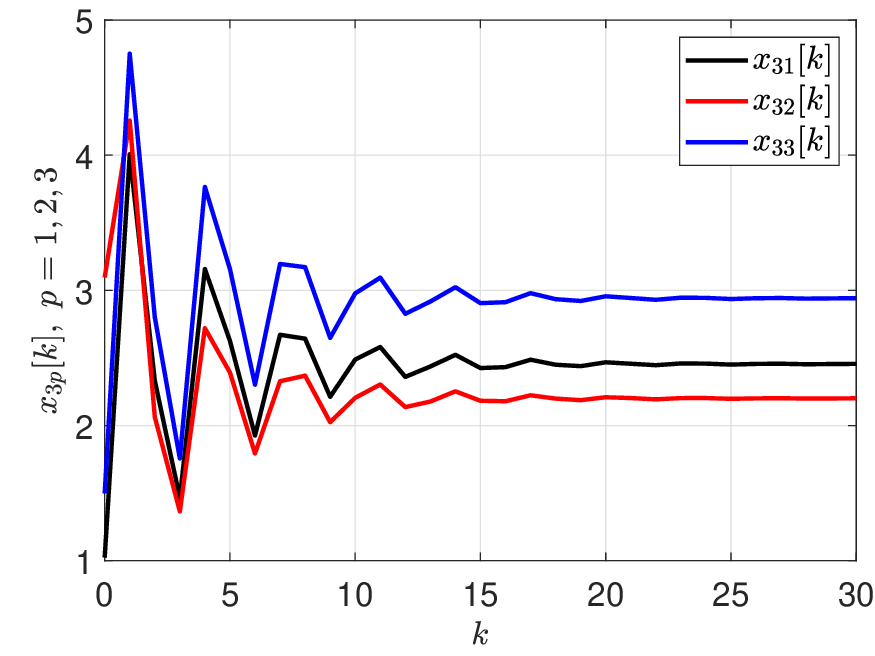}} \hfill
    \subfloat[]{\includegraphics[width=0.45 \linewidth]{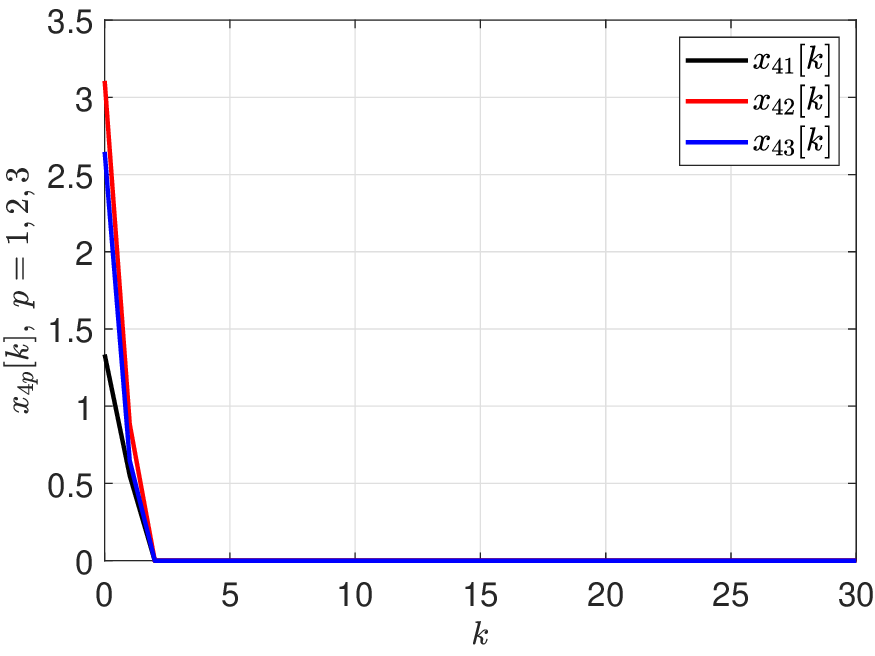}} \hfill
    \subfloat[]{\includegraphics[width=0.45 \linewidth]{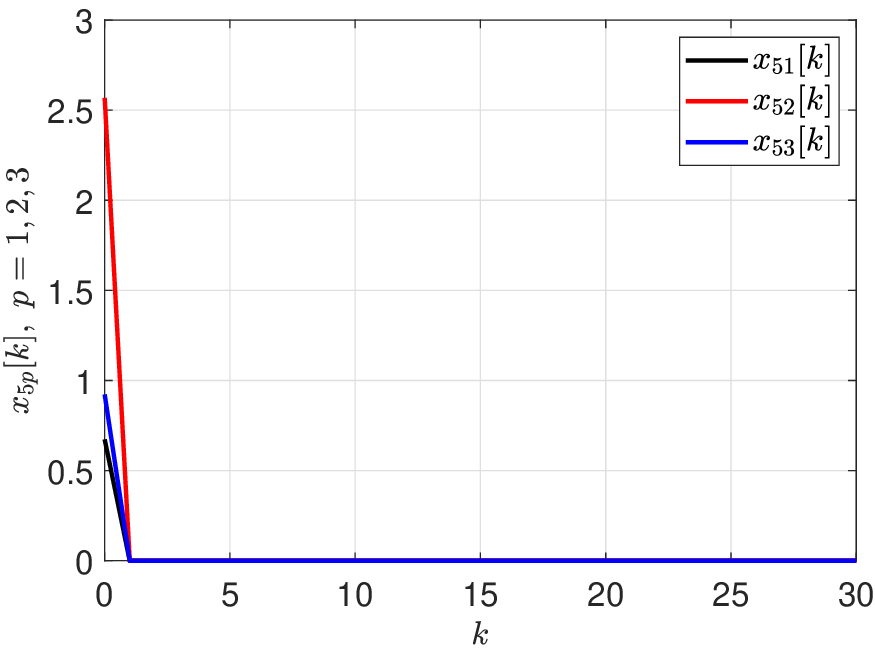}} \hfill
    \subfloat[]{\includegraphics[width=0.45 \linewidth]{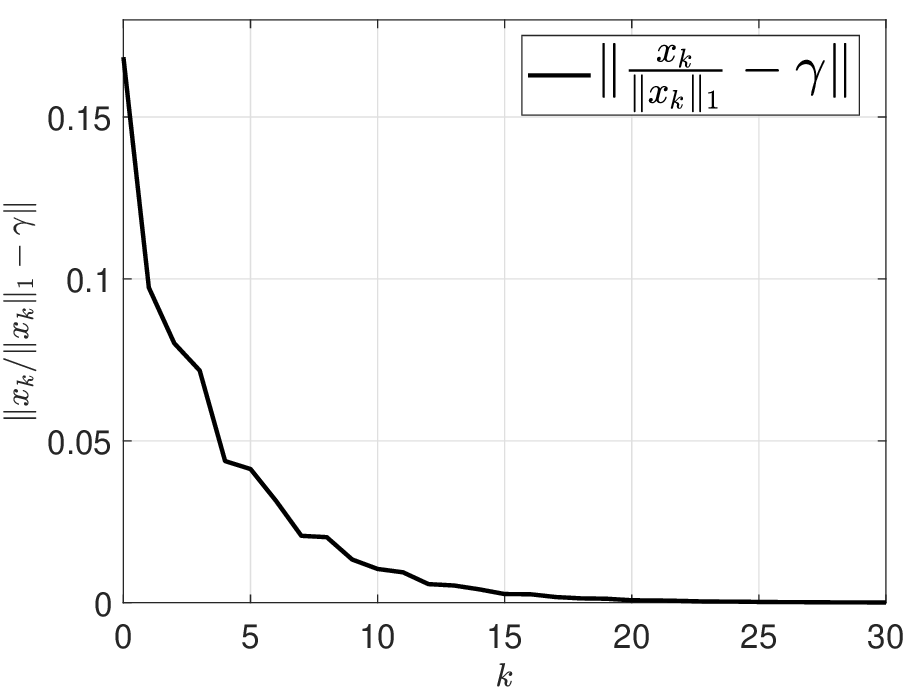}} \hfill
    \caption{(a)--(e): The states $x_{ip}[k]$, $i=1,\ldots,5,~p=1,2,3$ of 5 economic systems in the closed models. (g): The vector $x[k]/\|x[k]\|_1$ converges to the normalized right eigenvector $\bm{\gamma}$ corresponding to the unity eigenvalue of matrix $\bar{\m{A}}$. \label{fig:c15_closed_model}}
\end{figure}

\subsection{A closed input-output economic network}
In this section, the input-output matrices \[\m{A}_{11}= \begin{bmatrix}
    0.0 & 0.2 & 0.3\\0.5 & 0.6 & 0.5\\ 0.5 & 0.2 & 0.2
\end{bmatrix}, ~\m{A}_{ij} = \begin{bmatrix}
    0.6 & 0.2 & 0.2\\0.2 & 0.6 & 0.1\\ 0.2 & 0.2 & 0.7
\end{bmatrix},~ \] 
$\forall (v_i,v_j)\in \mc{E}\setminus (v_1,v_1),$ and the weights $w_{11}=0.4$, $w_{21}=0.6$, $w_{25}=w_{45}=0.5$, $w_{13}=w_{32}=w_{34}=1$ are considered. It can be checked that the network satisfies all assumptions of Theorem~\ref{thm:c15_closed_IO_model}.

Simulation results are depicted in Fig.~\ref{fig:c15_closed_model}. It can be seen that the output vectors $\m{x}_1,\m{x}_2,\m{x}_3$ asymptotically converge to some non-zero vectors while $\m{x}_4,\m{x}_5$ converge to $\m{0}_3$. Indeed, Fig.~\ref{fig:c15_closed_model}(g) shows that $\frac{\m{x}}{\|\m{x}\|_1}$ asymptotically converges to $\bm{\gamma}$ - the normalized right eigenvector corresponding to the unity eigenvalue of the matrix $\bar{\m{A}}$.

\subsection{An open input-output economic network}
\begin{figure}[t]
    \centering
    \subfloat[]{\includegraphics[width=0.45 \linewidth]{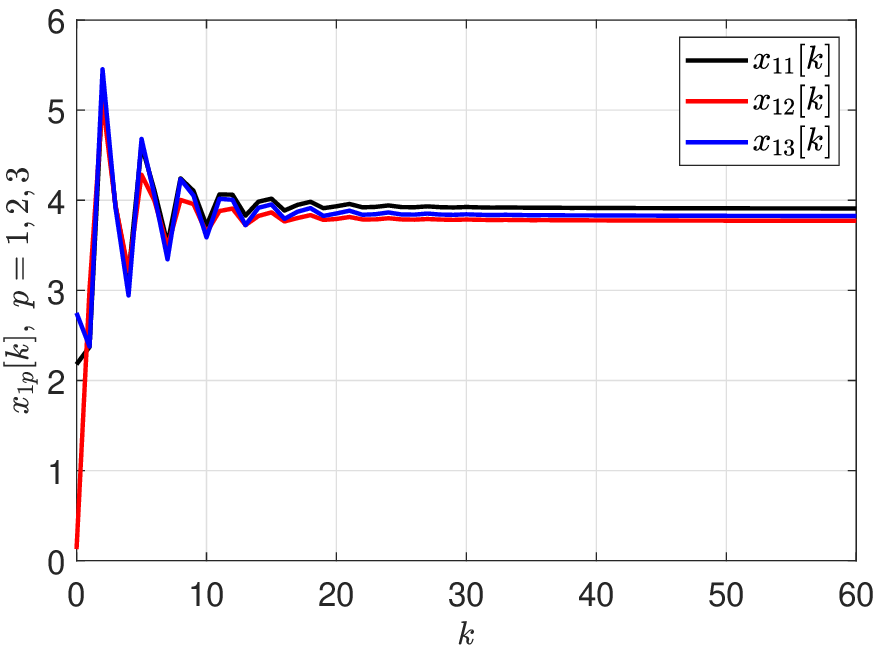}} \hfill
    \subfloat[]{\includegraphics[width=0.45 \linewidth]{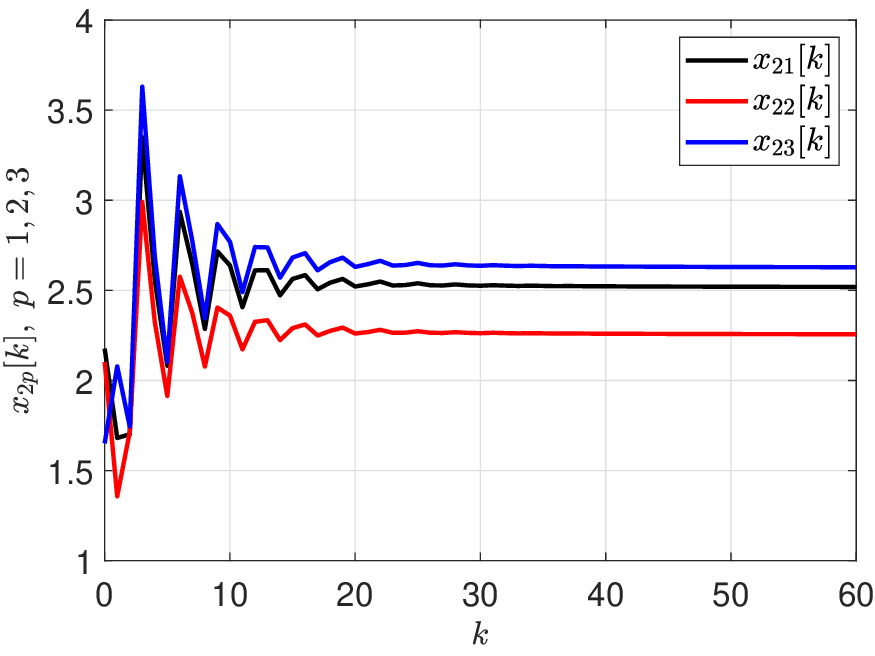}} \hfill
    \subfloat[]{\includegraphics[width=0.45 \linewidth]{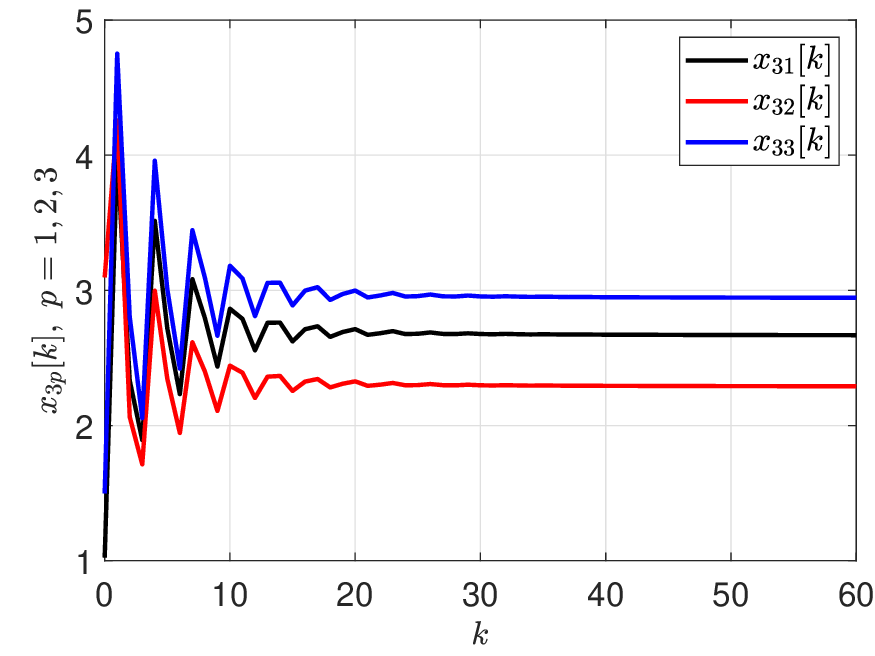}} \hfill
    \subfloat[]{\includegraphics[width=0.45 \linewidth]{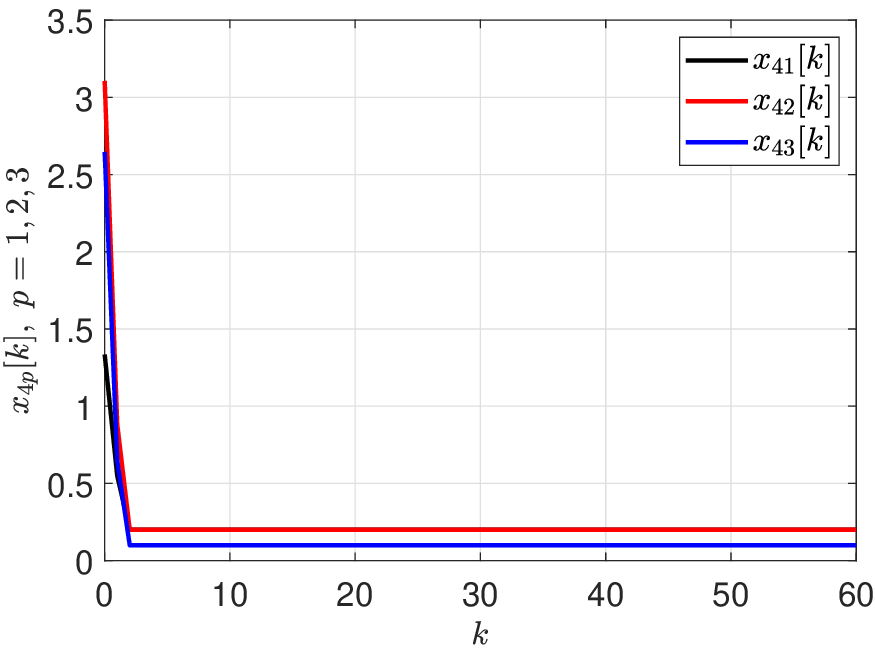}} \hfill
    \subfloat[]{\includegraphics[width=0.45 \linewidth]{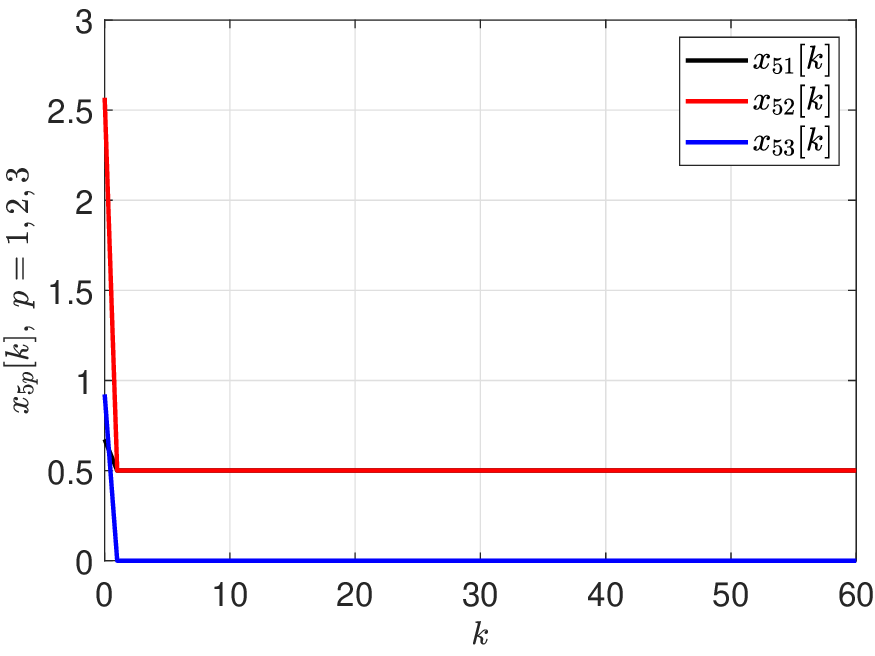}} 
    \caption{(a)--(e): The states $x_{ip}[k]$, $i=1,\ldots,5,~p=1,2,3$ of five economic systems in the open model. \label{fig:c15_open_model}}
\end{figure}
Consider the same network of five economic systems, in which the matrix $\m{A}_{11}$ is changed to 
\[\m{A}_{11} = \begin{bmatrix}
        0.0 & 0.2 & 0.3 \\
        0.2 & 0.4 & 0.1 \\
        0.2 & 0.0 & 0.3 \end{bmatrix}\]
It can be checked that all assumptions of Theorem~\ref{thm:c15_open-io-model} are satisfied. The demand vector is chosen as $0.5[{1}, 1,\m{0}_{10}^\top,1,1,0]^\top$, corresponding to the red vertices $L_1$--$L_4$ in Fig.~\ref{fig:c15_network}.

Simulation results depicted in Fig.~\ref{fig:c15_open_model} verify that $\m{x}[k]$ asymptotically converges to some equilibrium $\m{x}^*$. The existence of non-zero demand in the fifth economic system causes non-zero value of the equilibrium price in the fourth and the fifth economic systems. The demands $\m{y}_i$ act as leader agents (who do not change their state values over time) in a leader-follower matrix-weighted graph. Once the matrix $\m{I}-\bar{\m{A}}$ is invertible, the state vectors of the leaders uniquely determine an equilibrium price structure $\m{x}^*=(\m{I}-\bar{\m{A}})^{-1}\m{y}$. Specifically, in our network, the graph in Fig.~\ref{fig:c15_network} consists of five agents (15 vertices) and four leader vertices (L$1$--L$4$), the first- and the second industry of the fifth economic system have constant demands from L$3$ and L$4$, respectively. This causes $x_{51}[k]$, $x_{52}[k]$ to converge to 0.5 after a few iterations. In turn, $x_{51}[k]$, $x_{52}[k]$ act as nonzero demands for $x_{4p}[k]$, $p=1,2,3$, making them converge to non-zero final values. In contrast, as there is no demand on the third industry of the fifth economic system, $x_{53}[k]$ converges to 0 after a few iterations.

\section{Notes and references}
\label{sec:notes}
In this chapter, the networked input-output economic close/open models, which is built upon the classical works of W. Leontief and W. Issard and the theory of matrix-weighted graphs, were proposed. The proposed models have significantly generalized the celebrated input-output model and created a bridge between an economic problem and networked control. Distributed matrix-weighted iterative algorithms were proposed to determine the equilibrium price vector of the models. 

\section*{Acknowledgment}
The authors would like to thank Quoc Van Tran for his helpful comments related to this paper. The permission to use the material in this preprint in \cite[Chapter 15]{TrinhAhn2025MWG} was agreed by all authors.

\appendices
\bibliographystyle{IEEEtran}
\bibliography{minh2024}

\end{document}